\input{aipcheck.tex}

\edef\optionlist{%
   \variorefoptionifavailable        
   draft,%
   \psnfssproblemoption              
   tnotealph}

\documentclass[\optionlist]{aipproc}
\def\selectedlayoutstyle{8x11single}

\def\next{pdf}
\ifthenelse{\equal\selectedlayoutstyle{pdf}}
  {%
  \renewcommand\selectedlayoutstyle{8d}
   \usepackage[draft=false,bookmarksopen]
              {hyperref}
  }
  {}
\layoutstyle\selectedlayoutstyle

\received{x}
\accepted{x}

\usepackage{ttct0001}

\usepackage{shortvrb}
\MakeShortVerb\|

\makeatletter
   \def\@oddfoot{\reset@font
                 \copyright{} 2008 AIP
                 \hfil\@title
                 \hfil\@date\hfil\thepage}
\makeatother

\begin{document}

\author{Jihn E. Kim}{
  address={Department of Physics and Astronomy and Center for Theoretical Physics,\\ Seoul National University, Seoul 151-747, Korea},
  email={jekim@phyp.snu.ac.kr},
}


\def\Qem{{$Q_{\rm em}$}}
  \def\SMSSM{${\cal S}$MSSM\ }
\def\CPT{${\cal CPT}$\ }

 \def\N{{$\cal N$}}
 \def\Z{{\bf Z}}
 \def\MG{{$M_{\rm GUT}$}}

\def\EE{E$_8\times$E$_8^\prime$}

 \def\N{{$\cal N$}}
 \def\Z{{\bf Z}}
 \def\MG{{$M_{\rm GUT}$}}

\def\EE{E$_8\times$E$_8^\prime$}
\def\Eo{E$_8$}
\def\Eh{E$_8'$}
\def\MGUT{$M_{\rm GUT}$}
\def\ie{{\it i.e.}\ }

\def\fourb{\overline{\bf 4}}
\def\four{{\bf 4}}
\def\two{{\bf 2}}
\def\fiveb{\overline{\bf 5}}
\def\five{{\bf 5}}
\def\tenb{\overline{\bf 10}}
\def\ten{{\bf 10}}
\def\one{{\bf 1}}
\def\six{{\bf 6}}
\def\threeb{\overline{\bf 3}}
\def\three{{\bf 3}}

\title{String Compactification and Unification of Forces}
\date{2008/2/28}


\begin{abstract}
  I review our recent attempts toward obtaining
  the MSSM from string orbifold compactification. The required constraints are the existence of three families and R parity, vectorlike exotics, one pair of Higgs doublets, and the SU(5)$'$ hidden sector for dynamical breaking of SUSY toward a GMSB scenario. We also comment on the threshold correction which are influenced by a power law evolution of gauge couplings through the KK radius in non-prime orbifolds and can be used to fit the couplings.
\end{abstract}

\maketitle

\tableofcontents

\bigskip

\def\EE{{E$_8\times$E$_8'$}}

\section{Introduction}
This small workshop is on grand unification theories(GUT) and I will try to obtain GUTs or GUT-like standard models(SM) from string compactification. GUTs introduce the hierarchy problem and supersymmetry(SUSY) has been studied extensively in the last quarter century to understand the hierarchy problem. Now we are finally close to confronting experimental verification/falsification of TeV scale SUSY.
If superstring is relevant to low energy physics, it may reveal through an effective supergravity Lagrangian. So, the MSSM phenomenology is the first hurdle to overcome in string phenomenology.

String theory has been studied in many fronts. For obtaining the MSSM group, compactification of the \EE\ heterotic string has been most successful, and we follow this route in this talk. Let us start to glimpse the important issues in supergravity related low energy SUSY models:
\begin{itemize}
\item
In the last 24 years TeV SUSY has been based on supergravity Lagrangian given in \cite{Cremmer}.
\item
In supergravity, gravitino phenomenology is essential. One unavoidable constraint is the reheating temperature after inflation, $T_{rh}<10^{9-7}$ GeV \cite{EKN84,KKM05}.
\item
To verify the existence of gravitino, attempts to detect it at LHC has been proposed via the neutralino decay to
gravitino \cite{Buchmuller}.
\item
Most probably, we need an R parity for proton longevity.\footnote{However, a tiny violation of R parity can be tolerated \cite{Rviolation}.} In this regard, most existing string constructions are ruled out. Especially, the $u^cd^cd^c$ coupling must be forbidden.
\item
One has to solve the so-called $\mu$-problem \cite{muprob,GiuMas}. More generally, the MSSM problem, ``Why only one pair of Higgs doublets at the TeV scale?", must be understood.
\item
The strong CP problem must be resolved in the string framework, presumably by string axions \cite{staxion}.                 \item
 One has to resolve the SUSY flavor problem. The gauge mediated SUSY breaking (GMSB) exists in this regard \cite{GMSB}, and the recent surge of interest a la ISS \cite{ISS} reflects the seriousness of the SUSY flavor problem.
\item
 There exists the little hierarchy problem. At present the MSSM needs a fine tuning of order
 1\%, signaling 10-100 TeV SUSY particle masses.
 In this regard, the negative stop mass possibility has been considered to raise the fine tuning to the level of 5-10\% \cite{DermHDKim}.
 We hope that this little hierarchy problem will be understood in the end.
\item
 One has to understand the moduli stabilization. The KKLT scenario \cite{KKLT} led to the consideration of a phenomenologically interesting mirage mediation
 \cite{miragemed}.
\item
 It is required to allow only vectorlike exotics or is  better not to have any exotics.

\end{itemize}

Among these, here we single out the exotics problem which has not been emphasized widely.
Most string models accompany exotics. Chiral exotics are
dangerous phenomenologically. So, all exotics must be made vectorlike. In string construction, this is a nontrivial condition. Until recently, we did not find exotics-free models. But recently we find exotics-free models \cite{KimKyaeSM,GMSBst}, where however the weak mixing angle turn out to be not $\frac38$. We do not know whether there exist exotics free models with $\sin^2\theta_W=\frac38$. Except this weak mixing angle problem, in the exotics-free models the condition on singlet
VEVs is not so strong as in models with exotics, which is a great virtue.

This talk is a top-down approach, and if a specific example is considered then we cite $\Z_{12-I}$ orbifold models.

\section{R Parity and String Axions}

The R parity or matter parity in the MSSM is basically put in by hand: quarks and leptons are given an odd R parity, Higgs fields are given an even R parity.
Note that one of the merits of SO(10) GUTs is that it may have a good and reasonable R parity by assigning
\begin{eqnarray}
 &{\bf 16}: {\rm R= odd},\nonumber\\
 & {\bf 10}: {\rm R= even}.\nonumber
\end{eqnarray}
But we can understand this simply as the disparity between spinor (${\cal S}$) and vector (${\cal V}$)  representations as shown in Table \ref{tab:SOSV}.
\begin{table}[!h]
\begin{tabular}{lcccclccc}
\hline
\tablehead{1}{l}{b}{Weight} &\tablehead{1}{l}{b}{SU(3)$\times$ SU(2)} &\tablehead{1}{l}{b} {Notation} &\tablehead{1}{c}{b} {3(B-L)} &&\tablehead{1}{l}{b}{Weight} &\tablehead{1}{l}{b}{SU(3)$\times$ SU(2)} &\tablehead{1}{l}{b} {Notation} &\tablehead{1}{c}{b} {3(B-L)} \\
\hline
$(\underline{+~-~-}~\underline{+~-})$ & $(\three,\two)$ &$u_L,\ d_L$ & 1&$\quad\ \ $ &$(\underline{+~+~-}~{+~+})$ & $(\threeb,\one)$ &$u^c_L$ & $-1$\\[0.4em]
&&&& &$(\underline{+~+~-}~{-~-})$ & $(\threeb,\one)$ &$d^c_L$ & $-1$\\[0.4em]
$({+~+~+}~\underline{+~-})$  &$(\one,\two)$ &$\nu_{eL},\ e_L$ &$-3$& &$({-~-~-}~{+~+})$ & $(\one,\one)$ &$N^c_L$ & $3$\\[0.4em]
&&&& &$({-~-~-}~{-~-})$ & $(\one,\one)$ &$e^c_L$ & $3$
\\[0.2em]
\hline
$(\underline{-1~~-1~~0}~~0~~0)$&$(\three,\one)$&$D$&$4$& &$(\underline{1~~1~~0}~~0~~0)$&$(\threeb,\one)$&$D^c$&$-4$
\\[0.2em]
$({0~~0~~0}~~\underline{0~~-1})$&$(\one,\two)$&$H_u$&$0$& &$({0~~0~~0}~~\underline{1~~0})$&$(\one,\two)$&$H_d$&$0$
\\[0.2em]
\hline
\end{tabular}
\caption{ The spinor {\bf 16} and the vector \ten\ of SO(10). For the spinor, we choose odd numbers of minus signs. + and - denote $\frac12$ and $\frac{-1}{2}$, respectively, and  $B-L=(-2~-2~-2~0~0)$. The underline denotes permutations. {\bf 16} carries odd numbers of $B-L$ and {\bf 10} carries even numbers of $B-L$.}\label{tab:SOSV}
\end{table}

For example, ${\cal S}{\cal S}{\cal V}$
coupling is allowed, but ${\cal S}{\cal S}{\cal S}$  coupling is not allowed. This kind of disparity appears in the integer and half-integer angular momenta also in the SO(1,3) Lorentz group. Thus, $u^cu^cu^c$ is of the ${\cal S}{\cal S}{\cal S}$ type and it is forbidden at the cubic level. In this sense, E$_6$ is not good as a GUT because  SO(10) matter {\bf 16} and SO(10) Higgs {\bf 10} are put in the same {\bf 27} representation of E$_6$,
$$
  \bf                      27  = 16+10+1.
$$
Usually, in E$_6$ therefore one introduces extra ${\bf 27}$ and $\overline{\bf 27}$ just for Higgs, which do not mix with matter ${\bf 27}$.

However, this attractive feature is not automatically applicable in SO(10) theories with an ultraviolet completion. The reason is that there can exist the SO(10) singlets of the ${\cal S}$ type, e.g. $(-~-~-~-~-)$, and nonrenormalizable interactions allow $u^cd^cd^c\langle {\cal S}\rangle/M_P$, and the R parity problem reappears again. In the compactification of heterotic string, we note the E$_8$ adjoint representations appear with the types of
\begin{eqnarray}
 &{\cal S}: (++-+-+++),\cdots,\nonumber\\
 &{\cal V}: (1 -1 0 0 0 0 0 0),\cdots\nonumber
\end{eqnarray}
where the abbreviation $\pm$ denotes $\pm\frac12$.
Thus, in the heterotic string compactification the strategy is to put matter representations in ${\cal S}$ type and Higgs repsesentations in ${\cal V}$ type from the original E$_8$, and one must consider nonrenormalizable interactions also.

Toward this objective, the standard practice to obtain a discrete parity is to put it as a subgroup of an anomaly free U(1) gauge group. In string models, we can include the anomalous U(1) gauge group also since the anomaly is cancelled by the Green-Schwarz mechanism \cite{GreenSch}. Let us call this U(1) as U(1)$_\Gamma$. For example, consider a VEV of a scalar carrying an even $\Gamma$ charge, with $\Gamma=(2~2~2~0~0~0~0~0)$. If $P\in U(1)_\Gamma$, then $\Gamma =\rm odd\ integer\ for {\cal S}$ and $\Gamma =\rm even\ integer\ for {\cal V}$. Thus, $P$ is successfully embedded in $U(1)_\Gamma$. Yesterday, Mohapatra discussed {\bf 126} Higgs of SO(10) and the usefulness of {\bf 126} is simply because it belongs to the $\cal V$ type.

In the heterotic string, there are four possibilities for U(1)$_\Gamma$ by choosing odd number of 2s:
\begin{eqnarray}
\begin{array}{l}
B-L\propto (2~ 2~ 2~ 0~ 0~ 0~ 0),\\
X \propto (2~ 2~ 2~ 2~ 2~ 0~ 0~ 0)\ :\ X\ {\rm of\ the\ flipped\ SU(5)}\\
Q_1 = (0~0~0~0~0~2~0~ 0),\\
Q_2 = (0~0~0~0~0~0~2~ 0).
\end{array}
\end{eqnarray}

\section{The FCNC Problem in Supergravity Models and GMSB}

Our prime objective is obtaining the MSSM spectrum with no chiral exotics or even without exotics and at the same time implementing the R parity. Furthermore, requiring a successful hidden sector is very restrictive. There are very few such models if any, since I know only one model presented in this talk. If the hidden sector is introduced toward a dynamical symmetry breaking of SUSY, the best chance for the hidden sector is an SU(5)$'$ \cite{SU5ADS,GMSBun}.

The orbifold compactification is well known by now \cite{orbibook}. The \EE\ heterotic string gives a good gauge groups and string phenomenology is most successful here.
Our experience shows that any orbifold has a same order of complexity. For example, even though $\Z_3$ orbifold looks the simplest, actually the 27 fixed points makes it
very complicated. On the other hand, the $\Z_{12-I}$ looks very complicated, but it is simple in Wilson lines with only 3 fixed points and probably it is simpler than others if one knows how to construct models. In Table \ref{tab:Wilsonlinecond}, we list the conditions on Wilson lines. From this table, note that there are four cases of simple Wilson lines, which are underlined. Certainly, $\Z_3$ is simpler than $\Z_2$ on two-torus and hence $\Z_{6-I}$
and $\Z_{12-I}$  are simplest ones. Among these, only  $\Z_{12-I}$ are known to have phenomenologically interesting models \cite{Z12Iflip,SMZ12I,GMSBun,GMSBst}.

\begin{table}[!t]
\begin{tabular}{llll}
\hline
&\tablehead{1}{l}{b}{Lattice} &\tablehead{1}{l}{b} {Effective order} &\tablehead{1}{l}{b} {Conditions} \\
\hline
$\Z_3$ &SU(3)$^3$ &$3a_1=0,\ 3a_3=0,\ 3a_5=0$ & $a_1=a_2,\ a_3=a_4,\ a_5=a_6$\\[0.2em]
 $\Z_4$ &SU(4)$^2$ &$2a_1=0,\ 2a_4=0$ &$a_1=a_2=a_3,\ a_4=a_5=a_6$\\[0.2em]
 &SU(4)$\times$SO(5)$\times$SU(2) & $2a_1=0,\ 2a_5=0,\ 2a_6=0$  & $a_1=a_2=a_3,\ a_4=0$\\[0.2em]
 &SO(5)$^2\times$SU(2)$^2$ &$2a_2=0,\ 2a_4=0,\ 2a_5=0,\ 2a_6=0$& $a_1=a_3=0$\\[0.2em]
$\Z_{6-I}$ &SU(3)$\times$SU(3)$^2$ &$\underline{3a_1=0}$ & $a_1=a_2,\ a_3=a_4=a_5=a_6=0$ \\[0.2em]
$\Z_{6-II}$ &SU(2)$\times$SU(6) &$\underline{2a_1=0}$  & $a_2= a_3=a_4=a_5=a_6=0$  \\[0.2em]
 &SU(3)$\times$SO(8) &$3a_1=0,\ 2a_5=0$  & $a_1=a_2,\ a_3=a_4=0,\ a_5=a_6$  \\[0.2em]
 &SU(2)$^2\times$SU(3)$^2$ &$3a_1=0,\ 2a_3=0,\ 2a_4=0$  & $a_1=a_2,\ a_5=a_6=0$ \\[0.2em]
$\Z_{7}$ &SU(7) &$\underline{7a_1=0}$ & $a_1=a_2= a_3=a_4=a_5=a_6$  \\[0.2em]
$\Z_{8-I}$ &SO(8)$\times$SO(5) &$2a_1=0,\ 2a_6=0$  & $a_1=a_2= a_3=a_4,\ a_5=0$  \\[0.2em]
$\Z_{8-II}$ &SO(10)$\times$SU(2) &$2a_4=0,\ 2a_6=0$  &  $a_1=a_2= a_3=0,\ a_4=a_5$ \\[0.2em]
 &SU(2)$^2\times$SO(8) &$2a_1=0,\ 2a_5=0,\ 2a_6=0$  & $a_1=a_2= a_3=a_4$ \\[0.2em]
$\Z_{12-I}$ &E$_6$ &no restriction & $a_1=a_2= a_3=a_4=a_5=a_6=0$  \\[0.2em]
 &SU(3)$\times$SO(8) &$\underline{3a_3=0}$ & $a_3=a_4,\ a_1=a_2=a_5=a_6=0$  \\[0.2em]
$\Z_{12-II}$ &SU(2)$^2\times$SO(8) &$2a_1=0,\ 2a_2=0$  &  $ a_3=a_4=a_5=a_6=0$ \\[0.1em]
\hline
\end{tabular}
\caption{The string orbifolds and the Wilson line conditions.}\label{tab:Wilsonlinecond}
\end{table}

\begin{figure}[!h]
\resizebox{.5\columnwidth}{!}
{\includegraphics{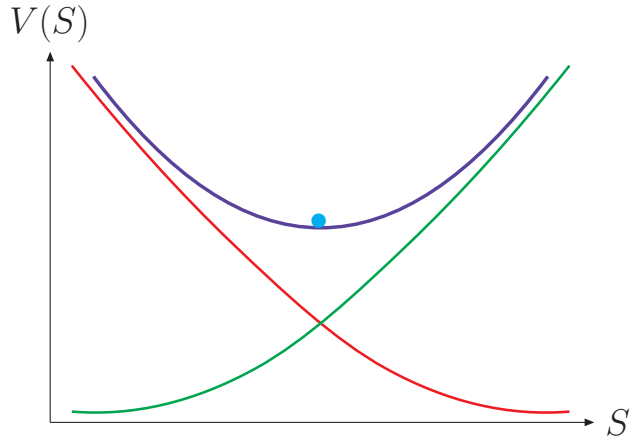}}
\caption{A potential shape for the dynamical SUSY symmetry breaking.}\label{fig:Abel}
\end{figure}

In supergravity models, there appear flavor changing neutral currents problems (FCNC) in general. Even if the superpotential is made flavor-conserving, the K\"ahler potential is restricted only by the reality, which is known to break the flavor symmetry. So, the SUSY flavor problem is generic in supergravity models. The SUSY flavor violations are parametrized by squark and slepton mixings $\delta_{LL}, \delta_{RR}, \delta_{LR}$ \cite{Masiero}, which are typically of ${\cal O}(10^{-2}\sim10^{ -3})$. This SUSY flavor problem led to the GMSB scenario \cite{DineNelson}. The well-known examples of dynamical SUSY breaking in simple groups are one family ($\ten$ plus $\fiveb$) SU(5) model \cite{SU5ADS} and ${\bf 16}+\ten$ of SO(10) \cite{SO10dyn}. These models are called uncalculable models \cite{dynSrev}. In this case, the behavior of vacuum energy is depicted in Fig. \ref{fig:Abel}.
In the figure, the runaway vacuum energy from the confining force \cite{VenYank} and the rising vacuum energy from a superpotential give a nonvanishing vacuum energy at a finite value of some fields.

\begin{figure}[!h]
\resizebox{.5\columnwidth}{!}
{\includegraphics{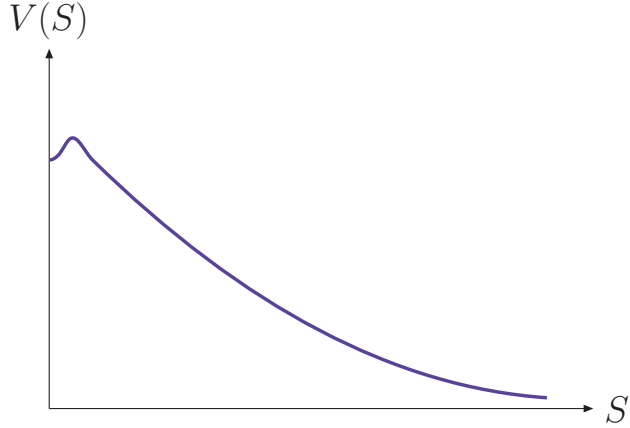}}
\caption{A potential shape for a local minimum at the origin of some field.}\label{fig:unstable}
\end{figure}

Because of the difficulty in obtaining one family SU(5) and ${\bf 16}+\ten$ of SO(10) in stable vacua, the recent study of unstable vacua suggested by Intrilligator, Seiberg and Shih (ISS) got a lot of interest \cite{ISSfollow} because it allows SUSY QCD with vectorlike quarks for dynamical SUSY breaking. Nelson and Seiberg argued for the need of R symmetry to break SUSY dynamically at the ground state \cite{NSRsymm}. ISS looked for a sufficiently long lived unstable vacuum, where the need of R-symmetry is discarded. In this case, the behavior of the potential is depicted in Fig. \ref{fig:unstable}.
Notably, SU(5)$'$ models with six or seven flavors allow SUSY breaking at an unstable vacuum at the origin of some fields \cite{ISS}.

In these GMSB models, messengers (symbolically denoted as $f$) of SUSY breaking to the observable sector are introduced. In the unstable vacuum models, a superpotential of the following form is introduced  \cite{Murayama06,ISSfollow},
\begin{equation}
W_{\rm tree}=m\overline{Q}Q +\frac{\lambda}{M_{\rm Pl}} \overline{Q}Q f\bar f +\bar ff,\quad {\rm for\ a\ local\ minimum}
\end{equation}
where $Q$ and $\overline{Q}$ is the hidden sector quark pair. The R symmetry breaking is introduced by the tree level
$W_{\rm tree}$, including the messengers $f$. Below the confining scale $\Lambda$, this superpotential can be discussed in terms of \cite{Murayama06},
\begin{equation}
W_{\rm ISS}=\overline{b}_i S^{ij}S_j-\frac{{\rm det}S^{ij}}{\Lambda^{N_f-3}}-m_i\Lambda S^{ii} \end{equation}
where the singlet $S$ develops an F-term and SUSY breaking is mediated by the messenger $f$ sector (generating F term by $W_{\rm tree}$) to the observable sector.

In the uncalculable models, the effective Lagrangian has a term of the form \cite{Murayama07,GMSBst},\footnote{We will comment more on this later.}
\begin{equation}
{\cal L}=\int d^2\theta\left(\frac{1}{M^2}W^{\prime\alpha}W'_\alpha\bar ff+ M_f\bar ff\right),\quad {\rm for\ stable\ vacuum}.
\end{equation}
In string models, there appear many heavy charged fields which can act as messengers \cite{GMSBst}.

To fulfil the condition for the DSB to occur
at a relatively low energy scale, later we will introduce different radii for the three comples tori. It is reminiscent of Horava and Witten's introduction of a distance between two branes in the M-theory \cite{Horava}. Also extra particles in the desert may be used to fit the data.

One attractive feature of SUSY GUTs is that with the desert hypothesis the coupling constants meet at $\alpha_{\rm GUT}\simeq \frac{1}{25}$ at the energy scale $(2-3)\times 10^{16}$ GeV. Because of the possibility of populating the desert between the TeV scale and the $(2-3)\times 10^{16}$ GeV scale, we may allow $\alpha_{\rm GUT}\simeq \frac{1}{20}-\frac{1}{30}$ at the unification point.
For the SUSY flavor problem in the GMSB scenario, the gravity mediation to soft parameters must be sub-dominant the GMSB contribution; thus we may requires the SUSY  breaking scale in the GMSB  scenario below $10^{11-12}$ GeV.

The GMSB scenario needs two ingredients:
\begin{itemize}
\item SUSY breaking sector in terms of a confining gauge group,
         e.g. SU(5)$'$, with hidden sector quarks $Q$ confining at the scale $\Lambda_h$.
\item Messengers of SUSY breaking at the scale $M_f$.
\end{itemize}
So, we consider the following scales
\begin{equation}
{\rm On\ } \Lambda_h:\ \frac{\Lambda_h^3}{M_P^2}\le 10^{-3}\ {\rm TeV}\Rightarrow \Lambda_h\le 2\times 10^{12}\ {\rm GeV}
\end{equation}
\begin{equation}
{\rm On\ a\ naive\ estimate\ of\ } M_f:\ \frac{\xi\Lambda_h^2}{M_f}\approx 10^3\ {\rm GeV}
\end{equation}
where $\xi$ is a model dependent number. Since $M_f< M_P$ is expected, $\Lambda_h$ may be smaller than $10^{12}$ GeV.
To estimate the confining scale of the hidden sector, $\Lambda_h$, we consider its one-loop coupling running
\begin{equation}
\frac{1}{\alpha^h{\rm GUT}}=\frac{1}{\alpha^h_j(\mu)} +\frac{-b_j^h}{2\pi}\ln\Big(\frac{M^h_{\rm GUT}}{\mu}\Big).
\end{equation}
If $b^h_j$ is given, $\Lambda_h$ can be calculated in terms of the inverse coupling $A'\equiv{1}/{\alpha^h_{\rm GUT}}$.
\begin{figure}[!t]
\resizebox{.6\columnwidth}{!}
{\includegraphics{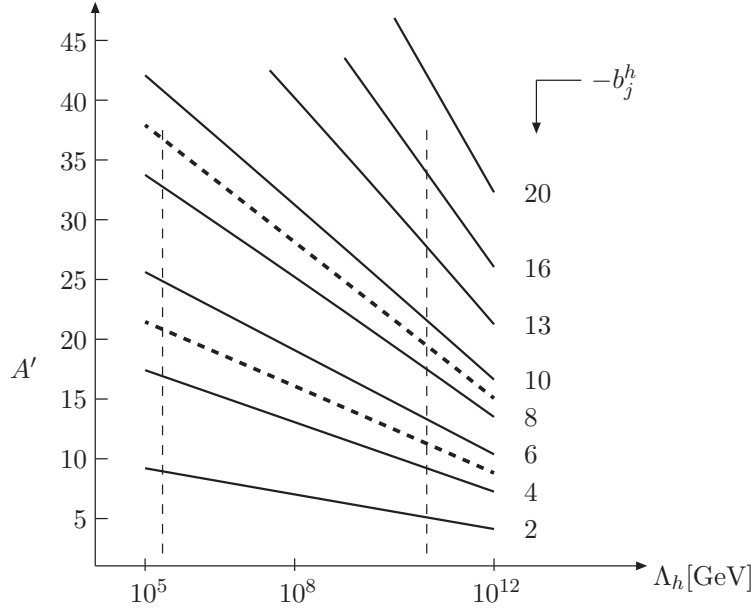}}
\caption{Constraints on $A'$. The confining scale is defined as the scale $\mu$ where $\alpha^h_j(\mu)=1$. Using $\xi=0.1, M_X=2\times 10^{16}$ GeV in the upper bound region and $\xi=0.1, M_X=\frac12\times 10^{6}$ GeV in the lower bound region, we obtain the region bounded by dashed vertical lines. Thick dash curves are for $-b_j^h=5$ and 9. }\label{fig:Hiddencoupl}
\end{figure}
The relation between $A'$ and $\Lambda_h$ is shown in Fig. \ref{fig:Hiddencoupl}. For example, we obtain $A'\simeq 27.4$ in SU(4)$'$ with no matter ($b^h_j=-12$). It may be difficult to find such a model anyway. For SU(5)$'$ with 7 flavors ($b^h_j=-8$) corresponding to an unstable vacuum \cite{ISS}, we obtain $A'\simeq 18.6$ and $\Lambda_h\sim 10^{9-10}$ GeV.

\section{Orbifolds with Kaluza-Klein Radius Dependence}

\begin{figure}[!]
\resizebox{.5\columnwidth}{!}
{\includegraphics{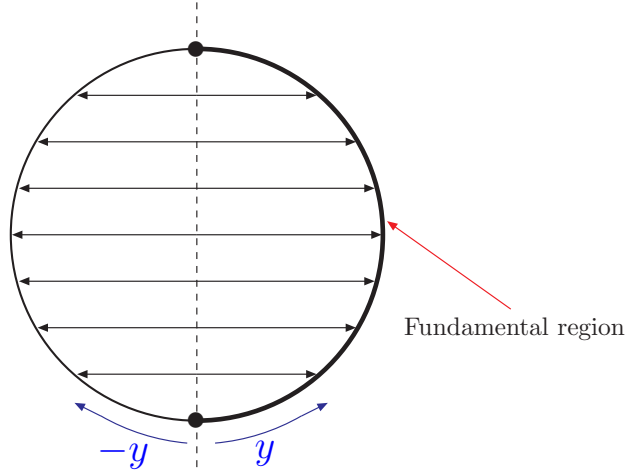}}
\caption{A shape of the orbifold $S_1/\Z_2$.}\label{fig:S1Z2}
\end{figure}

An orbifold is a manifold moded out by a discrete action. It was used extensively in the compactification of string models \cite{DHVW:1985}, and later adopted in extra-dimensional field theory \cite{kawamura}. The simplest example is 1-dimensional (1D) torus moded out by the $\Z_2$ discrete action as shown in Fig. \ref{fig:S1Z2}.
Because of the identification of two points by $\Z_2$, we can consider only the half set of the manifolds points except at the boundary. The boundary points are called fixed points and the region between the fixed points is called the fundamental region. The area of the fundamental region is the half of the area of the original manifold (circle). The discrete action, say ${\sf g}$ transforms the point in the manifold $z$ to ${\sf g}z$. This action ${\sf g}$ is an operator in quantum mechanics, and the wave functions are acted by this operator. In the $S_1/\Z_2$ orbifold action, the points in the manifolds transform
\begin{eqnarray}
{\sf g}: y\to -y
\end{eqnarray}
and a vector potential is acted as
\begin{eqnarray}
{\sf g}: \left\{
\begin{array}{l}
V_y(y)\to -V_y(-y)\\
V_\mu(y)\to +V_\mu(-y).
\end{array}\right.
\end{eqnarray}

\begin{figure}[!h]
\resizebox{.5\columnwidth}{!}
{\includegraphics{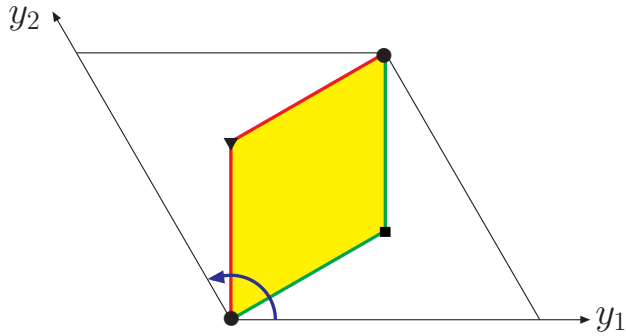}}
\caption{The $T_2/\Z_3$ orbifold with three fixed points $\bullet,\star,$ and $\times$. Its topology is a triangular ravioli.}\label{fig:T2Z3}
\end{figure}

Another simple orbifold used extensively in string models is a two-dimensional (2D) torus moded out by a $\Z_3$ action as shown in Fig. \ref{fig:T2Z3}.
In this $T_2/\Z_3$ orbifold, there are three fixed points and the area of the fundamental region is 1/3 (the yellow region of Fig. \ref{fig:T2Z3}) of the torus area.
The coordinate of the 2D torus is customarily represented by a complex number $z$, and the orbifold action is
\begin{eqnarray}
{\sf g}: z\to e^{2\pi i/3}z
\end{eqnarray}
and a vector potential is acted as
\begin{eqnarray}
{\sf g}:\left\{
\begin{array}{l}
V_z(z)\to e^{2\pi i/3}V_y(e^{2\pi i/3}z)\\
V_\mu(z)\to +V_\mu(e^{2\pi i/3}z).
\end{array}\right.
\end{eqnarray}

\subsubsection{A 5D SUSY GUT}

An interesting field theoretic orbifold is a 5D SUSY GUT with the internal space $S_1/\Z_2\times \Z'_2$. For the torus $S_1$, there are only two possibilities of discrete symmetries for moding out,
\begin{figure}[!h]
\resizebox{.47\columnwidth}{!}
{\includegraphics{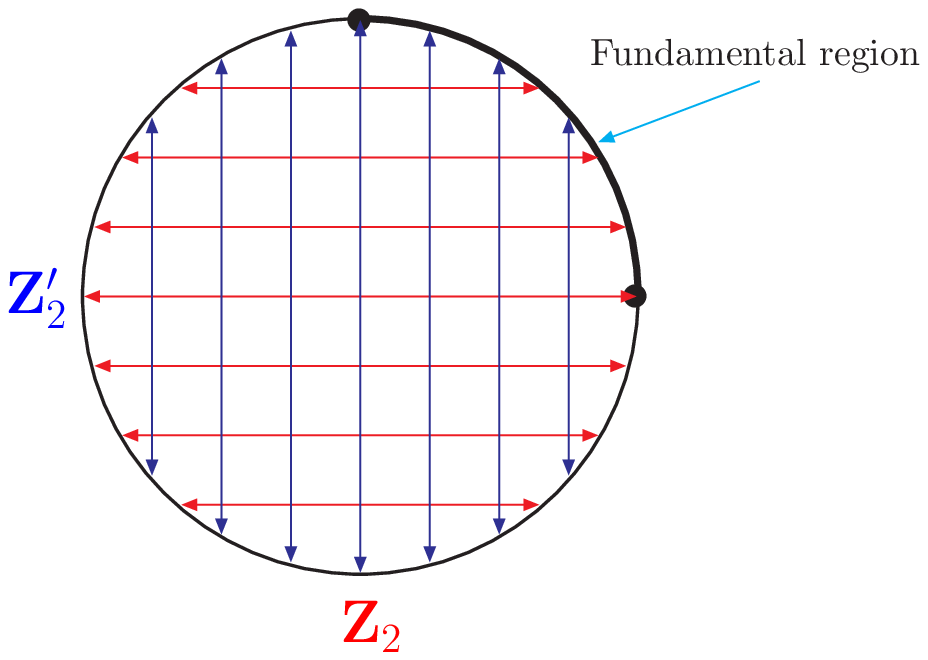}}
\caption{$S_1/\Z_2\times Z_2'$.}\label{fig:S1ZZ}
\end{figure}
one by $\Z_2$ and the other by $\Z_2\times \Z'_2$.
The  $S_1/\Z_2\times \Z'_2$ orbifold is shown in Fig. \ref{fig:S1ZZ}. As shown in Fig. \ref{fig:S1ZZ}, there are two fixed points and the area of the fundamental region is 1/4 of the area of the circle. The \N=1 5D SUSY has \N=2 in terms of 4D SUSY. An SU(5) GUT group is expected to be broken by the discrete action and also the \N=2 SUSY is also expected to be broken directly by the discrete action. Therefore, we need two $\Z_2$s as done in Ref. \cite{kawamura}. The 5D wave functions, with coordinate $(x^\mu,y)$ where $\mu=0,1,2,3$, have mode expansions in $\sum_n\phi(x)e^{iny}$ with mass $|n|/R$, which in other words has the $\cos ny$ and $\sin ny$ mode expansions. Thus, massless modes ($n=0$) appear only in the cosine mode or both the $\Z_2$ and $\Z_2'$ parities being +, $(Z_2,\Z_2')=(++)$. This method of obtaining massless modes is so simple and therefore attracted a great deal of attention.

\begin{figure}[!h]
\resizebox{.7\columnwidth}{!}
{\includegraphics{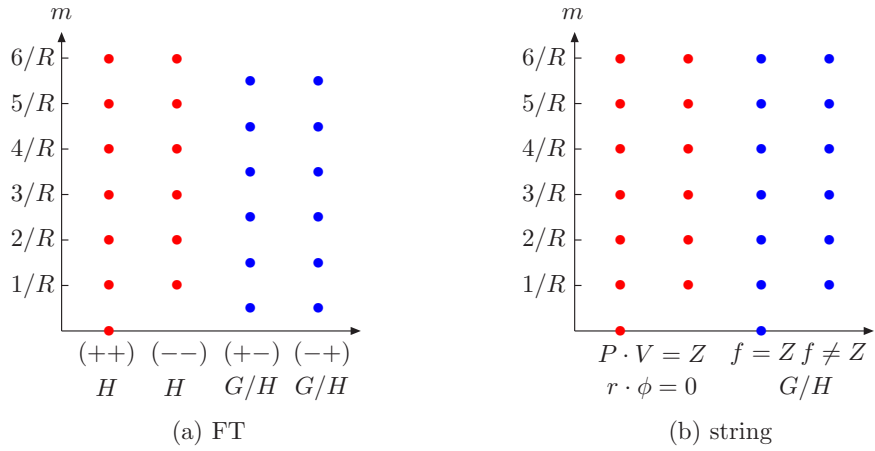}}
\caption{The KK tower in field theoretic and string orbifolds.}\label{fig:KKtower}
\end{figure}

In Fig. \ref{fig:KKtower}(a), we show the KK mass spectrum in field theoretic orbifld. To compare, in Fig. \ref{fig:KKtower}(b) we also show string theoretic \N=1 SUSY spectrum. Without SUSY, we note that one massless mode (++) is not paired by another as shown in  Fig. \ref{fig:KKtower}(a). On the other hand, string orbifolds with \N=1 SUSY, another SUSY partner appears as a massless mode also. Since the splitting of KK masses are $1/R$, we expect the spectra as shown in  Fig. \ref{fig:KKtower}.

In the SU(5) field theoretic orbifold $S_1/\Z_2\times \Z_2'$, the gauge multiplet ${\bf 24}_{SU(5)}$ splits into the SM representations with the following $(\Z_2,\Z_2')$ parities \cite{kawamura},
\begin{eqnarray}
{\bf 24}_{SU(5)}=\left(\begin{array}{cc}
({\bf 8},\one)_0^{(++)} & (\three,\two)_{-5/6}^{(+-)}\\
 (\threeb,\two)_{5/6}^{(+-)} & (\one,\three)_0^{(++)}
\end{array}
\right)\oplus
\left(\begin{array}{cc}
(\one)_{1/3}^{(++)} & 0\\
 0& (\one)_{-1/2}^{(++)}
\end{array}
\right)
\end{eqnarray}
The (++) states contain massless gauge boson modes, eight gluons plus four SU(2)$\times$U(1) gauge bosons as shown in Fig. \ref{fig:KKtower}(a). For the gauge multiplet, the \N=2 SUSY is broken down to \N=1 SUSY. The sector containing the SM gauge bosons and the sector $G/H$ (the so-called $X,Y$ gauge bosons) split as
\begin{eqnarray}
\left(\begin{array}{ccc}
A_\mu^{SM(++)} &\leftrightarrow &\lambda_2^{SM(--)}\\
  \lambda_1^{SM(++)} &\leftrightarrow &A_y^{SM(--)},\phi^{SM(--)}
\end{array}
\right),\quad
\left(\begin{array}{ccc}
A_\mu^{SU(5)/SM(+-)} &\leftrightarrow &\lambda_2^{SU(5)/SM(-+)}\\
  \lambda_1^{SU(5)/SM(+-)} &\leftrightarrow &A_y^{SU(5)/SM(-+)},\phi^{SU(5)/SM(-+)}
\end{array}
\right)
\end{eqnarray}
where the vertical transformation is the \N=1 SUSY transformation we are interested in and the horizontal transformation is the broken second \N$'$=1 SUSY transformation. In this model, two hypermultiplets $\five_H$ and  $\fiveb_H$ are introduced. Here, writing only the spin-0 components, the \N$'(\times$\N) relations between bosons are
\begin{eqnarray}
\five=\left(\begin{array}{c}
\three^{(+-)} \\
  \underline{H_u^{(++)}}
\end{array}\right)\leftrightarrow
\five^c=\left(\begin{array}{c}
\three^{c(-+)} \\
  H_u^{c(--)}
\end{array}
\right),\quad
\fiveb=\left(\begin{array}{c}
\threeb^{(+-)} \\
  \underline{H_d^{(++)}}
\end{array}\right)\leftrightarrow
\fiveb^c=\left(\begin{array}{c}
\threeb^{c(-+)} \\
  {H_d}^{c(--)}
\end{array}\right)
\end{eqnarray}
where the modes containing massless modes are underlined. Note that we obtain the doublet-triplet splitting in this model since one pair of Higgs doublets can be light while all colored Higgs fields are heavy. In orbifold string models, this possibility was noted long time ago \cite{IKNQ}.

With the above KK spectrum Dienes, Dudas and Ghegetta tried to calculate the evolution of gauge couplings above the TeV scale \cite{Dienes}. A typical form of the running of gauge couplings is
\begin{eqnarray}
\frac{1}{g^2_i(\mu)}\simeq \frac{1}{g^2_\Lambda} + b_i^0\ln \frac{\Lambda}{\mu}-(b_i^{(++)}+b_i^{(--)})
\ln\frac{\Lambda}{M_R}+\left(b_i^{(++)}+b_i^{(--)}+b_i^{(+-)}
+b_i^{(-+)}\right)
\left[\frac{\Lambda}{M_c}-1\right]
\end{eqnarray}
where $b_i^{(++)}+b_i^{(--)}$ comes from the \N=1 SM spectrum and $b_i^{(+-)}+b_i^{(-+)}$ comes from the SU(5)/SM sector. Here, $M_R=1/R$ and $M_c$ is the compactification scale. The contribution to  $b_i^{(++)}+b_i^{(--)}+ b_i^{(+-)}+b_i^{(-+)}$ is from the \N=2 SU(5) spectrum. The power dependence of the coupling constants appears in the last term.
 However, the field theoretic orbifold models are not ultraviolet completed models and the unification of coupling constant cannot be predicted. We will comment on the string threshold correction below. There, the KK spectrum follows the pattern given in Fig. \ref{fig:KKtower}(b) and in the $\Z_{12-I}$ it has the form \cite{Kyae07},
\begin{eqnarray}
\frac{16\pi^2}{g^2_H(\mu)}\simeq \frac{16\pi^2}{g_*^2} + b_H^0\ln \frac{M_*^2}{\mu^2}-\frac14 b_G^{N=2}
\ln\frac{M_*^2}{M_R^2}+\frac14 b_G^{N=2}
\left[\frac{2\pi}{\sqrt3}\frac{M_*^2}{M_c^2}-2.19\right].
\end{eqnarray}

\section{SU(5)$'$ Hidden Sector from $\Z_{12-I}$ Orbifold Compactification}

As discussed before, the most promising hidden sector group toward a GMSB is SU(5)$'$. If we want the hidden sector gaugino condensation, maybe there are more allowable choices restricted by $\Lambda_h\approx 10^{13}$ GeV only.
In our search of SU(5)$'$ hidden sector, we require
\begin{itemize}
\item Three chiral families,
\item SU(5)$'$ with $\ten+\fiveb$ or many pairs of $\five+\fiveb$,
\item Vectorlike exotics, or no exotics, which is another strong restriction.
\end{itemize}
Since we require three chiral families, it restricts very much the possible representations of the remaining gauge groups since in this orbifold compactifications the total number of chiral fields are not much more than 100. The obvious question is, $\lq\lq$Why $\Z_{12-I}$?" Probably,  $\Z_{12-I}$ is most restrictive in Yukawa couplings, and it has a simple Wilson line structure as discussed before \cite{Z12Iflip}. The restrictiveness is due to a large integer 12 used, and hence an approximate R-parity can be easily implemented \cite{RpApprox,KimKyaeSM}.

The $\Z_{12-I}$ twist is
\begin{equation}
\Z_{12-I}:\quad \phi=\left(\frac{5}{12}~\frac{4}{12}~\frac{1}{12} \right)
\end{equation}
where the second twist is $1/3$ which appears in the $\Z_3$ orbifolds.
\begin{figure}[!h]
\resizebox{.6\columnwidth}{!}
{\includegraphics{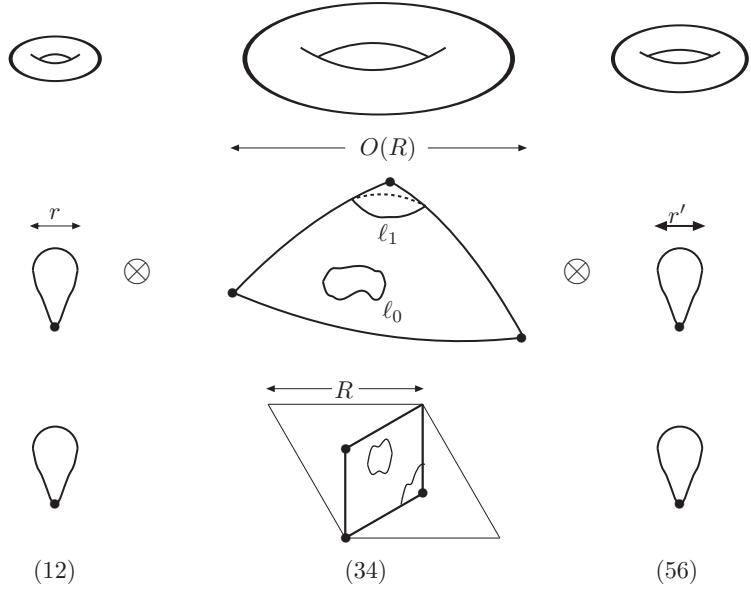}}
\caption{A schematic view of three torii in the $\Z_{12-I}$ orbifold.}\label{fig:threetorii}
\end{figure}
The shape of the $\Z_{12-I}$ orbifold is shown in Fig. \ref{fig:threetorii}, where the second orbifold has the shape of the $\Z_3$ orbifold. Here,
Wilson lines distinguish three fixed points. Only one (34)-torus and hence three fixed points of $\Z_{12-I}$ and 27 fixed points ($3^3$) in $\Z_3$. In the end, $\Z_3$ is as complicated as $\Z_{12_I}$ due to the complexity of Wilson lines. But, the geometric discussion is simpler in $\Z_{12_I}$ since we pay attention only to the (34)-torus. In $\Z_{12_I}$, much of breaking \EE\ is directly done by the shift vector $V$ only, which is the reason that the Wilson line $a_3$ can be simple.

In string compactification, the modular invariance conditions are to be satisfied. They correspond to choosing the $(++)$ parities and the anomaly cancelation conditions in field theory orbifold $S_1/{\Z_2\times\Z_2'}$. In $\Z_{12_I}$ the modular invariance conditions are
\begin{eqnarray}
\begin{array}{ll}
12(V^2-\phi^2)={\rm even\ integer}&\quad \\
12a_3^2={\rm even\ integer}&\quad \\
12V\cdot a_3={\rm  integer}&\quad
\end{array}
\end{eqnarray}
where $a_3=a_4$ and $a_1=a_2=a_5=a_6=0$.
The masslessness conditions are
\begin{eqnarray}
&&{\rm L-mover}:\ \frac{(P+kV)^2}{2}+ \sum_i{N^L_i} \tilde\phi_i-\tilde c_k=0\\
&&{\rm R-mover}:\ \frac{(\vec r+k\vec \phi)^2}{2}+ \sum_i{N^R_i} \tilde\phi_i-c_k=0
\end{eqnarray}
where $k=0(U),1,\cdots,11$. The generalized GSO projection calculates the multiplicity
\begin{eqnarray}
&&P_k(f)=\frac{1}{12\cdot 3} \sum_{l=0}^{N-1} \tilde\chi(\theta_k,\theta_l)e^{i2\pi \Theta_f}\\
&&\Theta_f=\sum_i (N_i^L-N_i^R)\hat\phi_i-\frac{k}{2}(V_f^2-\phi^2) + (P+kV_f)\cdot V_f - (\vec r+k\vec \phi)\cdot\vec\phi ,\quad
V_f=V+m_fa_3.
\end{eqnarray}

Let us briefly discuss two interesting $\Z_{12-I}$ models.

\subsection{A. $\Z_{12-I}$ model without exotics}
The Pati-Salam type gauge group but not exactly the same is obtained in Ref. \cite{GMSBun}. This model is commented briefly. The SU(4)$\times$SU(2)$_W\times $SU(2)$_V$ $\times$SU(5)$'$ representations are\footnote{Here, SU(2)$_V$ is not the same as the SU(2)$_R$ of the Pati-Salam model \cite{PatiS}.}
\begin{equation}
\begin{array}{ll}
U_1:~ (\fourb,\two,\one)_0,\ 2(\six,\one,\one)_0 &
 T_3:~  (\fourb,\one,\one)_{1/2},\ (\four,\one,\one)_{-1/2},\
  (\four,\one,\one)_{1/2},\ 2(\fourb,\one,\one)_{-1/2}\\
 U_2:~ 2(\four,\one,\two)_0,\ (\six,\one,\one)_0
 &\quad\quad 3(\one,\two,\one)_{1/2},\ 2(\one,\two,\one)_{-1/2},\
2(\one,\one,\two;\two;\one;\one)_{1/2}\\
\quad\quad  (\one,\two,\one;\one;\five';\one)_{-1/10},\
 2\cdot(\one,\two,\one;\one;\fiveb';\one)_{1/10}
 &\quad\quad  \
   (\one,\one,\two;\two;\one;\one)_{-1/2}\\
 U_3:~ (\four,\one,\two)_0,\ 2(\one,\two,\two)_0,\ (\one,\one,\one;\two;
 \one,\one)_0 &
T_{4_0}:~ 2(\one,\one,\one;\two;\one;\threeb')_0,\ 2\cdot\threeb'_0 \\
 T_{1_0}:~ (\fourb,\one,\one)_{1/2},\ (\one,\two,\one)_{1/2},
 \ (\one,\one,\two)_{1/2}&
 T_{4_+}:~ 2(\fourb,\two,\one)_0,\ 2(\four,\one,\two)_0,\
2(\six,\one,\one)_0,\ 7\cdot\two^n_0,
\ 9\cdot\one_0\\
T_{1_+}:~  (\one,\two,\one)_{-1/2},
 \ (\one,\one,\two)_{-1/2} &
 T_{4_-}:~ 2(\one,\one,\one;\two;\one;\three')_0,\ 2\cdot\three'_0\\
 T_{1_-}:~ (\one,\one,\two;\one;\five';\one)_{-1/10}
 & T_{7_+}:~ (\fourb,\one,\one)_{1/2},\ (\one,\one,\two)_{1/2}\\
 T_{2_0}:~  (\six,\one,\one)_{0},\
\two^n_0,\ \one_0 &
T_{7_-}:~ (\fourb,\one,\one)_{-1/2},\
(\one,\one,\two;\two;\one;\one)_{-1/2},\ (\one,\one,\two)_{-1/2} \\
 T_{2_+}:~ \five'_{2/5},\ \threeb'_0,&
  T_6:~ 6\cdot\fiveb'_{-2/5},\ 5\cdot\five'_{2/5},\  \\
T_{2_-}:~ (\one,\two,\two)_0,\ \three'_0,\ \two^n_0,\ 2\cdot\one_0&

\\
 \end{array}\label{Allspectrum}
 \end{equation}
where $\one=(\one,\one,\one;\one;\one;\one),\two^n=
(\one,\one,\one;\two;\one;\one),
\three'=(\one,\one,\one;\one;\one;\three')$ and
$\threeb'=(\one,\one,\one;\one;\one;\threeb')$.
We can see that the spectrum in (\ref{Allspectrum}) constitutes an anomaly free one. One attractive feature of this model is that it is free of exotics. In the hidden sector, we obtain SU(5)$'$ with the spectrum shown in Table \ref{tab:Hiddenun}.
\begin{table}[!h]
\begin{tabular}{ccc}
\hline \tablehead{1}{c}{b}{$P+n[V\pm a]$} &\tablehead{1}{c}{b} {Chirality} & \tablehead{1}{c}{b}{No.$\times$(Repts.)$_{Y,Q_1,Q_2}$
\tablenote{The 3rd and 4th row have SU(2)$_W$ doublets}} \\
\hline
$(\underline{\frac16~\frac16~\frac16~\frac{1}{3}~\frac{1}{3}~
 \frac{1}{12}~\frac{1}{4}
 ~\frac{1}{2}})
 (\underline{\frac34~\frac{-1}{4}~\frac{-1}{4}~\frac{-1}{4}
 ~\frac{-1}{4}}~
 \frac{-1}{4}~\frac{-1}{4}~\frac{-1}{4})'_{T1_-}$
 & $L$ & $
 ({\bf 1},{\bf 1},\two;1;{\bf 5}', 1)_{-1/10,-1/6,-4/3}^L$
\\[0.4em]
$(\frac{-1}{6}~\frac{-1}{6}~\frac{-1}{6}~~\frac{1}{6}~\frac{1}{6}~
 \frac{-1}{3}~0~\frac{1}{2} )
 (\underline{1~0~0~0 ~0}~
 0~0~0)'_{T2_+}$
  & $L$ & $({\bf 1},{\bf 1},{\bf 1};1;{\bf 5}', 1)_{2/5,-1/3,-8/3}^L$
\\[0.4em]
 $(0~0~0~\underline{\frac12~\frac{-1}{2}}~\frac{-1}{4}~\frac{1}{4}~0)
 (\underline{\frac34~\frac{-1}{4}~\frac{-1}{4}~\frac{-1}{4}
 ~\frac{-1}{4}}~\frac{1}{4}~\frac{1}{4}
 ~\frac{1}{4})'_{T3}$
  & $L$ & $({\bf 1},{\bf 2},{\bf 1};1;{\bf 5}', 1)_{-1/10,-1/2,0}^L$
\\[0.4em]
 $(0~0~0~\underline{\frac12~\frac{-1}{2}}~\frac{1}{4}~\frac{-1}{4}~0)
 (\underline{\frac{-3}{4}~\frac{1}{4}~\frac{1}{4}~\frac{1}{4}
 ~\frac{1}{4}}~\frac{-1}{4}~\frac{-1}{4}
 ~\frac{-1}{4})'_{T9}$ & $L$ & $
 2({\bf 1},{\bf 2},{\bf 1};1;\overline{\bf 5}', 1)_{1/10,1/2,0}^L$
\\[0.4em]
$(0~0~0~0~0~\frac{-1}{2}~\frac{1}{2}~0)
 (\underline{\textstyle -1~0~0~0~0}~0~0~0)'_{T6}$ & $L$ & $
 4({\bf 1},{\bf 1},{\bf 1};1;\overline{\bf 5}', 1)_{-2/5,-1,0}^L$
\\[0.4em]
$(0~0~0~0~0~\frac{-1}{2}~\frac{1}{2}~0)
 (\underline{\textstyle 1~0~0~0~0}~0~0~0)'_{T6}$ & $L$ & $
 2({\bf 1},{\bf 1},{\bf 1};1;{\bf 5}', 1)_{2/5,-1,0}^L$
\\[0.4em]
$(0~0~0~0~0~\frac{1}{2}~\frac{-1}{2}~0)
 (\underline{\textstyle -1~0~0~0~0}~0~0~0)'_{T6}$ & $L$ & $
 2({\bf 1},{\bf 1},{\bf 1};1;\overline{\bf 5}', 1)_{-2/5,1,0}^L$
\\[0.4em]
$(0~0~0~0~0~\frac{1}{2}~\frac{-1}{2}~0)
 (\underline{\textstyle 1~0~0~0~0}~0~0~0)'_{T6}$ & $L$ & $
 3({\bf 1},{\bf 1},{\bf 1};1;{\bf 5}', 1)_{2/5,1,0}^L$
\\[0.2em]
\hline
\end{tabular}
\caption{Hidden sector SU(5)$'$ representations. We picked up the left-handed chirality only from $T_1$ to $T_{11}$ representations.}
\label{tab:Hiddenun}
\end{table}
The hidden sector has ten pairs of $\five'$ and $\fiveb'$. So, by making 3 or 4 flavors of SU(5)$'$ heavy by the Higgs mechanism, we obtain the GMSB scenario at the unstable vacuum \cite{ISS}. However, this model is not attractive in that the hidden sector quarks carry the SM quantum number(s), in particular there are SU(2)$_W$ doublet hidden sector quarks.

So, the $\theta^0$ component VEVs of $\five-\fiveb$ condensate mesons is almost zero and SU(2)$_W$ is not broken at the SUSY breaking scale by $\theta^0$ component VEVs. But $\theta^2$ components are large and carry SU(2)$_W$ quantum numbers. So, our model, even though very attractive, is breaking SM at the SUSY breaking scale (by the meson F-term and baryon VEVs) and not working as a realistic model.

\subsection{B. Another exotics free model at stable vacuum}

The model presented in \cite{GMSBst} is very interesting in realizing
\begin{itemize}
\item Three chiral families
\item No exotics
\item Realization of R parity
\item One pair of Higgs doublets
\item The GMSB at a stable vacuum.
\end{itemize}
But the compactification scale value of the weak mixing angle ($\sin^2\theta_W$) is not $\frac38$, and it remains to be seen whether it renormalizes correctly to the observed one at the electroweak scale. The model is
\begin{equation}
\begin{array}{l}
V =\frac{1}{12}(6~6~6~2~2~2~3~3)(3~3~3~3~3~1~ 1~ 1)¡¯\\
a_3=\frac{1}{12}(1~1~2~0~0~0~0~0)(0~0~0~0~0~1~1 -2)¡¯
\end{array}
\end{equation}
Gauge group is SU(3)$_c \times$SU(3)$_W \times$SU(5)$'\times $ SU(3)$'\times$U(1)s.
which contains the Lee-Weinberg electroweak model \cite{LW77}. The model has no exotics. The observable sector fields are shown in Table \ref{tab:SMFamily}.
 \begin{table}[h]
\begin{tabular}{cccc}
\hline  \tablehead{1}{c}{b}{$P+[kV+ka]$}  &
\tablehead{1}{c}{b} {No.$\times$(Repts.)$_{Y[Q_1,Q_2,Q_3,Q_4,Q_5]}$}
 &\tablehead{1}{c}{b} {$\Gamma$} &\tablehead{1}{c}{b}{Label}\\
\hline
 $(\underline{\frac{-1}{3}~\frac{-1}{3}~\frac{-2}{3}}~
 \underline{\frac{2}{3}~\frac{-1}{3}}
 ~\frac{-1}{3}~0~0)(0^8)'_{T_{4_-}}$
  & $
 3\cdot(\three,\two)_{1/6~[0,0,0;0,0]}^L$
  & $1$ &$ q_1,~ q_2,~  q_3$
\\[0.4em]
$(\underline{\frac16~\frac{1}{6}~\frac{5}{6}}~
 {\frac{1}{6}~\frac{1}{6}}
 ~\frac{1}{6}~\frac12~\frac12)(0^8)'_{T_{4_-}}$   &$
 2\cdot(\threeb,\one)_{-2/3~[-3,3,2;0,0]}^L$
  & $3$ &$ u^c,~  c^c$
\\[0.4em]
$(\underline{\frac{-1}{3}~\frac{-1}{3}~\frac{-2}{3}}~
 {\frac{1}{3}~\frac{1}{3}}~\frac{1}{3}
 ~\frac{-1}{4}~\frac{-1}{4})(\frac14 ^5~\frac{1}{12}~\frac{1}{12}
 ~\frac{1}{12})'_{T_{7_+}}$   &$
 (\threeb,\one)_{-2/3~[0,6,-1;5,1]}^L$
  & $1$ &$ t^c$
\\[0.4em]
$(\underline{\frac{1}{2}~\frac{1}{2}~\frac{1}{2}}~
 {\frac{-1}{6}~\frac{-1}{6}}~\frac{-1}{6}
 ~0~0)(0^5~\frac{-1}{3}~\frac{-1}{3}
 ~\frac{-1}{3})'_{T_{2_0}}$   &$
 (\threeb,\one)_{1/3~[3,-3,0;0,-4]}^L$
 & $-1$ &$ d^c$
\\[0.4em]
$(\underline{\frac16~\frac{1}{6}~\frac{5}{6}}~
 {\frac{1}{6}~\frac{1}{6}}
 ~\frac{1}{6}~\frac{-1}2~\frac{-1}2)(0^8)'_{T_{4_-}}$  &$
 2\cdot(\threeb,\one)_{1/3~[-3,3,-2;0,0]}^L$
 & $1$ &$ s^c,~ b^c$
\\
[0.4em]\hline
 $({\frac{-1}{3}~\frac{-1}{3}~\frac{1}{3}}~
 \underline{\frac{2}{3}~\frac{-1}{3}}~\frac{2}{3}
 ~0~0)(0^8)'_{T_{4_-}}$  &$
 (\one,\two)_{-1/2~[-6,6,0;0,0]}^L$
  & $1$ &$l_1,l_2,l_3$
\\
[0.4em]\hline
 $(0~0~0~\underline{\frac{2}{3}~\frac{-1}{3}}~
 \frac{2}{3}~\frac{-1}{4}~\frac{-1}{4})
 (\frac14 ^5~\frac{1}{12}~\frac{1}{12}
 ~\frac{1}{12})'_{T_{1_0}}$  &$
 (\one,\two)_{1/2~[0,6,-1;5,1]}^L$
  & $0$ &$H_u$
\\[0.4em]
$(\frac{-1}{3}~\frac{-1}{3}~\frac{1}{3}~
\underline{\frac{1}{3}~\frac{-2}{3}}~\frac{1}{3}~
  ~\frac{-1}{4}~\frac{-1}{4})(\frac14 ^5~\frac{1}{12}~\frac{1}{12}
 ~\frac{1}{12})'_{T_{7_+}}$  &$
 (\one,\two)_{-1/2~[-6,0,-1;5,1]}^L$
  & $-2$ &$H_d$
\\
[0.4em]\hline
\end{tabular}
\caption{Three families of quarks and leptons and a pair of Higgs doublets. We do not list singlet leptons since there are many possibilities.} \label{tab:SMFamily}
\end{table}
Note that U(1)$_\Gamma$ charges of the SM fermions are odd and those of the Higgs doublets are even. Therefore, by breaking $U(1)_\Gamma$  by VEVs of even $\Gamma$ singlets, we break U(1)$_\Gamma$ to a discrete matter parity $P$ or $R$ parity. Thus, we achieve realizing a successful $R$ parity \cite{GMSBst}. Extra vectorlike doublets are given superheavy masses.

Because of the $R$ parity, the dimension-4 coupling $u^cd^cd^c$ coupling is not present. The dimension-5 coupling of the form $qqql$ is not forbidden by the $R$ parity, but in the present model it is forbidden up to a very high order because of the remaining U(1) gauge symmetries which are listed as subscripts in Table \ref{tab:SMFamily}.

\subsubsection*{The $\mu$ problem and one pair of Higgs doublets}

\begin{figure}[!h]
\resizebox{.7\columnwidth}{!}
{\includegraphics{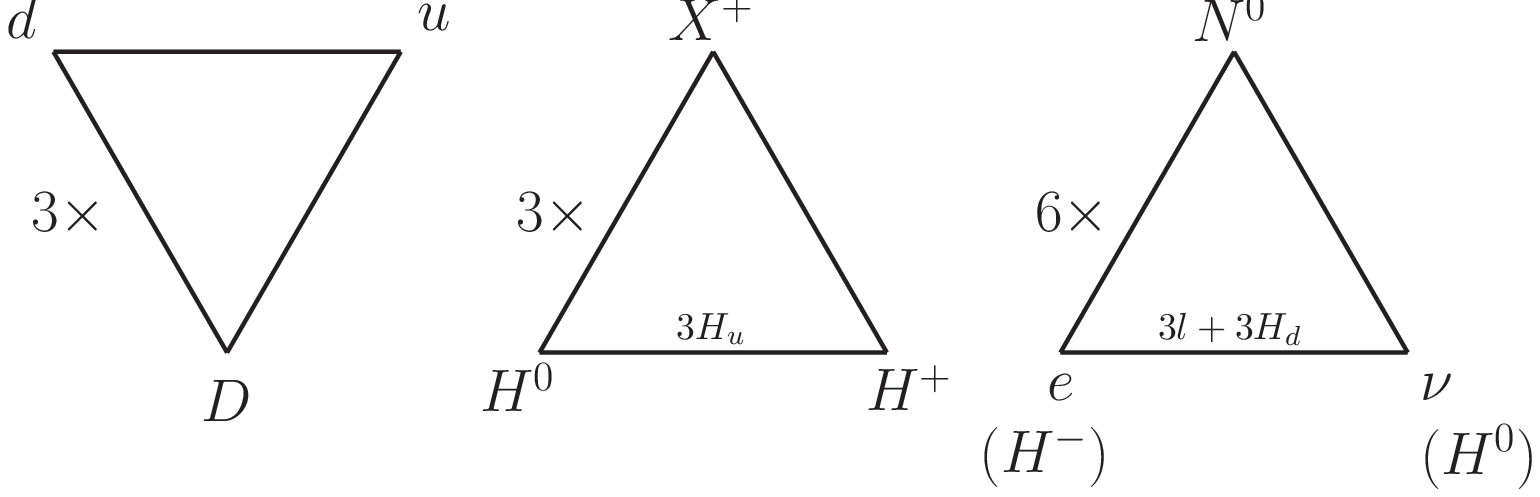}}
\caption{The {\bf 3} and {$\threeb$} representations of the Lee-Weinberg model.}\label{fig:LeeWein}
\end{figure}

Except the three chiral families, the remaining representations form a vector-like one. Generally, if not forbidden by a special symmetry, vector-like representations including Higgs doublets are heavy. Thus, the need for one pair of Higgs doublets is difficult to realize in general, which is the so-called $\mu$ problem. In our model, we present a novel mechanism for allowing one light pair of Higgs doublets. It is achieved because the electroweak gauge group is the Lee-Weinberg SU(3)$_W\times$U(1). From Table \ref{tab:SMFamily}, there appear three quark weak triplets, which appear in three colors and hence count in total 9 weak triplets from the quark sector, $3(\three_c,\three_W)$. From the anomaly cancelation, at low energy therefore we have 9 color singlet weak anti-triplets. This situation is shown in Fig. \ref{fig:LeeWein}. These color singlet weak triplets are split, according to their quantum numbers, into $\threeb_W(H^+),\threeb_W(H^-),$ and $\threeb_W(\rm lepton)$.
Three $\threeb_W(\rm lepton)$ remain light because of the chirality. The remaining representation $3[\threeb_W(H^+)+\threeb_W(H^-)]$ is vectorlike after the breaking of SU(3)$_W\times$U(1) down to SU(2)$_W\times$U(1)$_Y$. However, we can consider the original SU(3)$_W\times$U(1) for the discussion of Yukawa couplings.

Note that in our model both $H^+$ and $H^-$ appear from $\threeb$. It is in contrast to the other
cases such as in SU(5) SUSY GUT or SO(10) SUSY GUT. Therefore, in our case SU(3)$_W$ invariant $H^+$ and $H^-$  coupling must come from $\threeb_W\wedge\threeb_W\wedge\threeb_W$. Thus, there appears the Levi-Civita symbol and two $\epsilon$ symbols must be introduced, one from taking the SU(3)$_W$ singlet and the other from the flavor basis! (What else?) Therefore, in the flavor space the $H^+$ and $H^-$ mass matrix must be
antisymmetric and hence its determinant is zero, we conclude there appear one pair of massless Higgs doublets.
Thus, the MSSM problem of Ref. \cite{Kim03} is resolved.

It is interesting to compare our result to the old
introduction of color:
\begin{itemize}
 \item  In 1960s, it was known that the low-lying baryon and meson multiplets are embedded in {\bf 56} of the old SU(6) which is completely symmetric.
     But spin-half quarks are better to be fermions, which led to the introduction of an
     antisymmetric index. That was the famous SU(3)$_c$ of color, describing strong interactions \cite{HanNambu}.
\item In our supersymmetric theory, superpotential is described by bosonic fields, i.e. by the $\theta^0$ components of chiral multiplets.
 Since the Lee-Weinberg SU(3)$_W$ introduces an antisymmetric index, we need to introduce another antisymmetric flavor index \cite{GMSBst}.
\end{itemize}

\subsubsection*{Hidden sector}
The hidden sector gauge group for breaking SUSY is SU(5)$'$ and in Table \ref{tab:Hiddenstable} we list the SU(5)$'$ representations.
 \begin{table}[t]
\begin{tabular}{ccc}
\hline  \tablehead{1}{c}{b}{$P+n[V\pm a]$} &  \tablehead{1}{c}{b}{$~\Gamma~$} &
\tablehead{1}{c}{b}{ No.$\times$(Repts.)$_{Y[Q_1,Q_2,Q_3,Q_4,Q_5]}$}
\\[0.2em]
\hline $({\frac{1}6~\frac{1}6~\frac{-1}6~\frac{1}{6}~\frac{1}{6}~
 \frac{1}{6}~\frac{1}{4}~\frac{1}{4}})
 (\underline{\frac{-3}4~\frac{1}{4}~\frac{1}{4}~\frac{1}{4}
 ~\frac{1}{4}}~
 \frac{-1}{4}~\frac{-1}{4}~\frac{-1}{4})'_{T1_-}$
 & $2$ & $
 (\one;\fiveb', \one)_{0~[3,3,1;1,-1]}^L$
\\[0.4em]
$(\frac{1}{6}~\frac{1}{6}~\frac{-1}{6}~
\frac{-1}{6}~\frac{-1}{6}~
 \frac{-1}{6}~0~0)
 (\underline{\frac{1}{2}~\frac{1}{2}~\frac{-1}{2}~\frac{-1}{2}
 ~\frac{-1}{2}~}\frac{-1}{6}~\frac{-1}{6}
 ~\frac{-1}{6}~)'_{T2_+}$
 & $-1$ & $\star~({\bf 1};\ten', \one)_{0~[3,-3,0;-2,-2]}^L$
\\[0.4em]
 $(0^6~\underline{\frac14~\frac{-3}{4}})
 (\underline{\frac34~\frac{-1}{4}~\frac{-1}{4}~\frac{-1}{4}
 ~\frac{-1}{4}}~\frac{1}{4}~\frac{1}{4}
 ~\frac{1}{4})'_{T3}$
  & $-1$ & $(\two_n;{\bf 5}',\one)_{0~[0,0,-1;-1,3]}^L$
\\[0.4em]
  $(0^6~\underline{\frac34~\frac{-1}{4}})
 (\underline{\frac{-3}4~\frac{1}{4}~\frac{1}{4}~\frac{1}{4}
 ~\frac{1}{4}}~\frac{-1}{4}~\frac{-1}{4}
 ~\frac{-1}{4})'_{T9}$
  & $1$ & $(\two_n;\fiveb',\one)_{0~[0,0,1;1,-3]}^L$
  \\[0.4em]
 $(0^3~{\frac{-1}{3}~\frac{-1}{3}}~\frac{-1}{3}
 ~\frac{1}{4}~\frac{1}{4})
 (\underline{\frac{-3}4~\frac{1}{4}~\frac{1}{4}~\frac{1}{4}
 ~\frac{1}{4}}~\frac{1}{12}~\frac{1}{12}
 ~\frac{1}{12})'_{T7_0}$
  & $-1$ & $\star~(\one;\fiveb',\one)_{0~[0,-6,1;1,1]}^L$
\\[0.4em]
 $(\frac{1}6~\frac{1}6~\frac{-1}6~
 \frac{1}6~\frac{1}6~\frac{1}6~\frac{-1}{4}~\frac{-1}{4})
 (\underline{\frac{3}4~\frac{-1}{4}~\frac{-1}{4}~\frac{-1}{4}
 ~\frac{-1}{4}}~\frac{1}{4}~\frac{1}{4}~\frac{1}{4})'_{T7_-}$
  & $0$ & $(\one;\five',\one)_{0~[3,3,-1;-1,3]}^L$
\\[0.4em]
$(0^6~\frac{-1}{2}~\frac{-1}{2})
 (\underline{\textstyle -1~0~0~0~0}~0~0~0)'_{T6}$ & $-2$ & $
 3\cdot(\one;\fiveb', \one)_{0~[0,0,-2;-4,0]}^L$
\\[0.4em]
$(0^6~\frac{-1}{2}~\frac{-1}{2})
 (\underline{\textstyle 1~0~0~0~0}~0~0~0)'_{T6}$ & $-2$ & $
 2\cdot(\one;{\bf 5}', \one)_{1~[0,0,-2;4,0]}^L$
\\[0.4em]
$(0^6~\frac{1}{2}~\frac{1}{2})
 (\underline{\textstyle -1~0~0~0~0}~0~0~0)'_{T6}$ & $2$ & $
 2\cdot(\one;\fiveb', \one)_{-1~[0,0,2;-4,0]}^L$
\\[0.4em]
$(0^6~\frac{1}{2}~\frac{1}{2})
 (\underline{\textstyle 1~0~0~0~0}~0~0~0)'_{T6}$ & $2$ & $
 3\cdot(\one;\five,\one)_{0~[0,0,2;4,0]}^L$
\\[0.2em]
\hline
\end{tabular}
\caption{Hidden sector SU(5)$'$  representations under
SU(2)$_n\times $SU(5)$'\times $SU(3)$'$. After removing vectorlike representations by $\Gamma=$ even integer singlets, the starred representations remain.} \label{tab:Hiddenstable}
\end{table}
After removing the vectorlike representation there remain $\ten'$ and $\fiveb'$ which are starred in Table \ref{tab:Hiddenstable}.
This set of chiral representations is the source of dynamical SUSY breaking in SU(5)$'$ \cite{ADS}, and an F term appears for the chiral gauge multiplet, ${\cal W}'^\alpha{\cal W}_\alpha'$ for example \cite{Murayama07}. This F term splits the SUSY partner masses of messengers $f$,
\begin{equation}
{\cal L}=\int d^2\theta \left(\frac{1}{M^2}\bar ff{\cal W}'^\alpha{\cal W}_\alpha'+M_f \bar ff\right)+{\rm h.c.}
\label{GMSBmessint}
\end{equation}
where $M$ is the parameter, presumably above 10$^{12}$ GeV. For example, vectorlike $Q_{\rm em}=-\frac13$ $D$-type quarks  can be colored messengers. The superpartner mass splittings of the messenger sector transmit the information to the observable sector via gauge interactions and hence the soft masses of squarks and sleptons appear as flavor independent \cite{DineNelson}.
\vskip 0.5cm

\noindent $\Big[${\bf Noted added:}

One can see that the ${\cal W}'{\cal W}'$ in Eq. (\ref{GMSBmessint}) develops an F term. We can consider the following operators below the confinement scale of SU(5)$'$,
\begin{eqnarray}
&&Z\sim {\cal W}^{a}_b{\cal W}^b_{a} \label{Zuncalc}\\
&&Z'\sim \epsilon_{acfgh}{\cal W}^{a}_b{\cal W}^c_{d} \ten^{eb}\fiveb_e \ten^{fd}\ten^{gh}\label{Zpuncalc}
\end{eqnarray}
where the contraction of spinor indices of the chiral gauge multiplet ${\cal W}$ is implied. Under the global symmetry
U(1)$_A\times$U(1)$_B\times$U(1)$_R$, $\ten, \fiveb$ and ${\cal W}'$ transform as
\begin{eqnarray}
\begin{array}{cccc}
& U(1)_A & U(1)_B & U(1)_R\\
\ten & p & 1 & r\\
\fiveb & q &-3 & s\\
{\cal W}& 0 & 0 & 1\\
Z & 0&0& 2\\
Z' &3p+q& 0&3r+s+2
\end{array}
\end{eqnarray}
where U(1)$_B$ charges are given as anomaly free. U(1)$_R$ is also chosen as anomaly free, which gives the relation $3r+s=-6$. Thus, $Z'$ carries $R=-4$. On the other hand U(1)$_A$ is anomalous. The fermionic zero modes contributes to the instanton amplitudes as $e^{-8\pi^2/g^2(\mu)+i\theta}=(\Lambda/\mu)^{3N_c-\sum_f \ell(f)}= (\Lambda/\mu)^{15-2}$ where $\Lambda$ is the dynamically generated mass scale. The so-called 't Hooft's determinental instanton amplitude carries flavors (e.g. represented as $2N_c$ gluino lines plus $2N_f$ quark lines in SUSY QCD) and hence after integrating out the one-loop beta function we assign $2\sum_f \ell(f)$ charge for the U(1)$_A$ quantum number to the scale $\Lambda^{(3N_c-\sum_f \ell(f))}$.\footnote{See, for example, Ref. \cite{IntSeib85}.}  Thus, $\Lambda^{13}$ carries the U(1)$_A$ charge $3p+q$. Including this instanton amplitude, we try to include all possible terms allowed by U(1)$_A\times$U(1)$_B\times$U(1)$_R$. Namely, U(1)$_A$ for the instanton interaction is respected if we consider the combination $\Lambda^{13}$ divided by $\ten\cdot\ten\cdot\ten\cdot\fiveb$,\footnote{With one $\ten$ and one $\fiveb$, the combination $\ten\cdot\ten\cdot\ten\cdot\fiveb$ is not possible, but $Z'$ is possible as shown in (\ref{Zpuncalc}). However, other combinations of ${\cal W}'{\cal W}'$ with matter fields are not possible.} or $\Lambda^{16}$ by $Z'$. In this way, we can write all possible terms.  U(1)$_A\times$U(1)$_R$ symmetries dictate the following effective superpotential, after redefining $Z$ and $Z'$ as dimension-1 fields,
\begin{eqnarray}
W_{\rm eff}=\sum_a c_a m^{2-3a}Z^{1+2a}Z'^{a}\label{EffW}
\end{eqnarray}
where $c_a$ are dimensionless constants. The determinental interaction corresponds to $a=-1$. Strong dynamics may also allow the $a=0$ term. In (\ref{EffW}) we included all terms allowed just from the symmetry argument. Considering only the two terms with $a_{-1}$ and $a_0$ the SUSY conditions cannot be satisfied simultaneously, but a runaway solution results. So, we consider at least three terms, for which we choose $a=-1, 0$ and 1, for an illustration. Then, the SUSY conditions are
\begin{eqnarray}
&&\frac{\partial W}{\partial Z}=-c_{-1}m^5 Z^{-2}Z'^{-1}+ c_0m^2+3c_1m^{-1}Z^2Z'=0\\
&&\frac{\partial W}{\partial Z'}=-c_{-1}m^5Z^{-1}Z'^{-2}+ c_1m^{-1}Z^3=0
\end{eqnarray}
which cannot be satisfied simultaneously unless $c_0+2\sqrt{c_1c_{-1}}=0$. The symmetry principle allows many terms including the determinental interaction, and in general SUSY is broken.
$\Big]$

\subsubsection*{Discrete symmetries}

\begin{figure}[!h]
\resizebox{.5\columnwidth}{!}
{\includegraphics{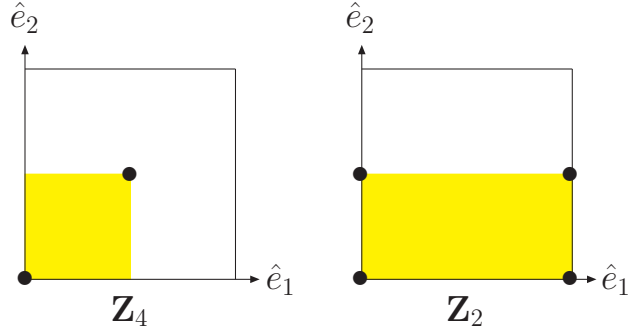}}
\caption{The degeneracy of fixed points due to the absence of Wilson lines.}\label{fig:fpdegeneracy}
\end{figure}

In the nonprime orbifolds, there are invariant torii in which case there exist some discrete symmetries. These discrete symmetries can be used for obtaining fermion mass spectrum. In the $\Z_{12-I}$ orbifold, $T_3$ and $T_6$ sectors have the following twists
\begin{equation}
\Z_{12-I}:\quad 3\phi=\left(\frac14~0~\frac14\right) ,\quad 6\phi=\left(\frac12~0~\frac12\right)
\end{equation}
which are $\Z_4$ and $\Z_2$, respectively. Thus, they have the fixed points as shown in Fig. \ref{fig:fpdegeneracy}.
For example, in $T_6$ we may consider an $S_4$ symmetry because the four fixed points cannot be distinguished by Wilson lines. The Yukawa couplings must respect this kind of discrete symmetry, which can be used to obtain nonabelian discrete symmetries by a further manipulation \cite{Koba}.

\section{Threshold correction}

The threshold correction via one loop is the torus topology, and the orbifolds on torus is the natural place to consider one loop corrections of closed strings. In string compactification, the threshold correction comes from non-prime orbifolds. The reason is that they contain invariant torus, and a large radius $R$ can be introduced. In $R\to \infty$, we have a 6D model, i.e. we obtain an \N=2 SUSY. The \N=2 models are vectorlike in 4D, and have masses of the form $1/R$ times integer. The simplest invariant torus is the $\Z_3$ substructure.

\begin{figure}[!h]
\resizebox{.7\columnwidth}{!}
{\includegraphics{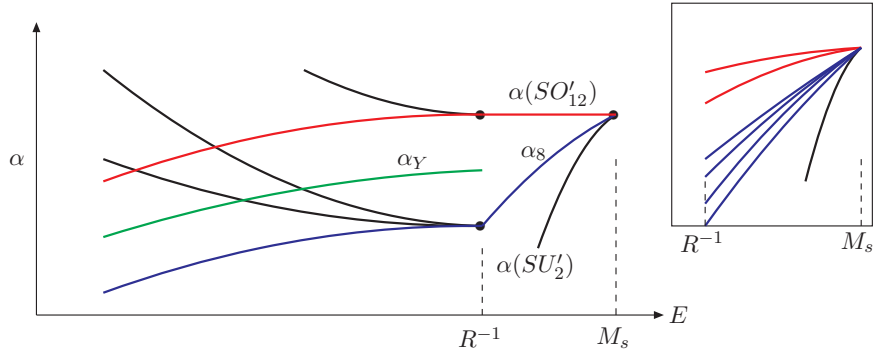}}
\caption{A schematic view of gauge coupling evolution without KK modes correction. $\alpha_8$ is the 6D SU(8) coupling and the green line for $\alpha_Y$ is the hypercharge coupling in Model S. The KK modes split couplings above $R^{-1}$ as depicted within the square.}\label{fig:couplevolution}
\end{figure}

The pioneering work on the threshold correction in string models has been calculated in Ref. \cite{Kaplunovsky,Dixon:1990pc,Antoniadis}, and we have recently implemented the method to add Wilson lines \cite{Kyae07}. The invariant sublattices are under $G'\in G$. Here, the \N=2 SUSY KK masses are described by a large radius ($R$), encoded in modulus of the metric. The simplest substructure $\Z_3$ appears in $\Z_{6-I}$ and $\Z_{12-I}$ orbifolds. But, there has not appeared a  phenomenologically interesting $\Z_{6-I}$ model, and we restrict the discussion to the $\Z_{12-I}$ models. Specifically, we work with the model presented in Ref. \cite{KimKyaeSM}:
\begin{eqnarray}
\begin{array}{l}
\phi=(\frac{5}{12}~\frac{4}{12}~\frac{1}{12})\\
\\
V=(\frac14~\frac14~\frac14~\frac14~\frac14~ \frac{5}{12}~\frac{5}{12}~\frac{1}{12}) (\frac14~\frac34~0~0~0~0~0~0)\\
\\
a_3=(\frac{2}{3}~\frac{2}{3}~ \frac{2}{3}~\frac{-2}{3}~
\frac{-2}{3}~\frac{2}{3}~0~\frac{2}{3}~) (0~\frac{2}{3}~\frac{2}{3}~0^5)
\end{array}
\end{eqnarray}
The invariant torus is the second one, i.e. the (34)-torus which obeys the $\Z_3$ identification. The radius $R$ of the (34)-torus is large compared to the compactification radii of (12)- and (56)-torii. Introduction of the Wilson line $a_3$ in the (34)-torus breaks $G$ down to the SM gauge group.  The anticipated evolution of gauge couplings are shown in Fig. \ref{fig:couplevolution}.

In contrast to our calculation, in an extra-dimensional field theory one cannot calculate the constant and $R^2$ term reliably.

We can obtain 6D field theory by  compactifying 4 internal
spaces, which is another check of our partition function
approach. Indeed these two calculations agree on the spectrum.

Integration in the modular space along the
above formula {\it a la} Dixon, Kaplunovsky and Louis \cite{Dixon:1990pc}
\begin{equation}
\Delta_i=\frac{Z'}{Z}\ b_i^{N=2} \int_\Gamma \frac{d^2\tau}{\tau_2}\left(\hat Z_{\rm torus}(\tau,\bar\tau)-1\right)\label{DKLformula}
\end{equation}
where $Z'=3$ (from the (34)-torus) and $Z=12$ in our case. Eq. (\ref{DKLformula}) gives the compactification size ($R$) dependence through the modular parameter with the following metric,
\begin{eqnarray}
&\left\{ \begin{array}{l}
\hat e_1=(\sqrt2,0) \\
\hat e_2=(-\sqrt{1/2},\sqrt{3/2})
\end{array}
\right. ;\quad
g_{ab}=\left( \begin{array}{cc}
2&-1 \\
-1& 2
\end{array}
\right)\\
&\left\{ \begin{array}{l}
\hat e^{*1}=(\sqrt{1/2},\sqrt{1/6}) \\
\hat e^{*2}=(0,\sqrt{2/3})
\end{array}
\right. ;\quad
g^{ab}=\frac13\left( \begin{array}{cc}
~2~&~1~ \\
1& 2
\end{array}
\right)
\end{eqnarray}
Thus, we obtain the following $R$ dependence of the gauge couplings,
\begin{equation}
\frac{4\pi}{\alpha_{H_0}(\mu)}
=\frac{4\pi}{\alpha_*}+b_{{ H}_0}^0~{\rm
log}\frac{M_*^2}{\mu^2}-\frac{b_{ H}}{4}\left[{\rm
log}\frac{R^2}{\alpha'}+1.89\right]+\frac{(b_{ H}+b_{
G/H})}{4} \left[\frac{2\pi R^2}{\sqrt{3}\alpha'}-0.30\right]
\label{gaugerun}
\end{equation}
where $H_0$ is the SM gauge group. Between $R$ and the string scale, the contribution to the $\beta$-function coefficient is given by $b_H$:
the corresponding group may not be the SM group. The $\beta$-function coefficient $b_H+b_{G/H}$ is the coefficient for the full group $G$ which is the gauge group obtained by $V$. Introduction of the Wilson line $a_3$ in the (34)-torus introduces the $R$ dependence. Because of the string calculation in our scheme, the resultant power behavior is reliable.
It is a reliable calculation, not like the expressions written in extra dimensional field theory \cite{Dienes}.
In particular, we point out that the
R-squared and constant terms are also reliable, and predicts how gauge couplings behave above the so-called GUT scale.

Between $R$ and the string scale, the contribution
to beta function coefficient is given by $b_H$. We note that
the corresponding group may not be the SM group.
Actually, we need singlet Higgs VEVs to give large masses for exotic particles \cite{KimKyaeSM}. Since SU(4) above the scale $1/R$ gives a complicated form for its U(1) subgroup, we break the SU(4) by VEVs of these singlets. So, we consider only the subgroup SU(2)$_W$ of the broken SU(4) and consider the N=2 $b_i$ (the $b_H$ term in Eq. (\ref{gaugerun})) in terms of another parameter $h_i$,
\begin{equation}
b_i=h_i\left(\log\frac{M_s^2}{M_R^2} +1.89\right).
\end{equation}

The hypercharge definition must be made
judiciously to avoid chiral exotics or even
to remove all exotics, as discussed in \cite{KimKyaeSM}: Model E with vectorlike exotics and Model S without exotics,
\begin{eqnarray}
&&{\rm Model\ E}:\ Y_E=(\frac{1}{3}~\frac{1}{3}~\frac{1}{3}~\frac{-1}{2} ~\frac{-1}{2}~0~0~0)(0^8)',\quad\sin^2\theta_W=\frac38\\
&&{\rm Model\ S}:\ Y_S=(\frac{1}{3}~\frac{1}{3}~\frac{1}{3}~\frac{-1}{2} ~\frac{-1}{2}~0~0~0)(0~0~1~0^5)',\quad
\sin^2\theta_W=\frac3{14}.
\end{eqnarray}
In Model E, $\sin^2\theta_W$ is the standard one and we obtain the usual result. Here, however, there exists another parameter $R$ which can be used to fit the strong coupling constant $\alpha_s(M_Z)$ to the observed value \cite{KimKyaeSM}. On the other hand, Model S has a much smaller $\sin^2\theta_W$ and the parameters $R$ and the string scale $M_s$ can be used to fit to the observed values of the mixing angle and the strong coupling, $\sin^2\theta_W(M_Z)= 0.22306\pm 0.00033$ and $\alpha_s(M_Z)=0.1216\pm 0.0017$ \cite{PData}. The allowed regions of $R=M_R^{-1}$ and $M_s$ are
\begin{equation}
{\rm Model\ S}:\quad \frac{M_R}{M_Z}\approx 1.70\times 10^{15},\ \ \frac{M_s}{M_R}\approx 3.68.
\end{equation}

\section{Conclusion}

We showed some interesting explanations of the SUGRA problems by the orbifold compactification the \EE\ heterotic string. We also considered the GMSB possibility in the orbifold compactification with a desirable MSSM spectrum. We observed that a 6D SUSY GUT is realized with the KK mass dependent threshold corrections. These corrections are reliable unlike in extra-dimensional field theory.  In some models,
three families appear with no exotics.  The GMSB at a stable vacuum in SU(5)$'$ with $\ten'$ plus $\fiveb'$ is shown to be possible. In this model, we obtained just the MSSM spectrum, i.e. with one pair of Higgs doublets. The R parity embedding is shown to be successful. We also discussed the gauge coupling unification in nonprime orbifolds with the KK mode contribution to the evolution equation. The KK mass parameter $1/R$ is used to obtain the coupling unification even with a GUT scale value $\sin^2\theta^{\rm GUT}_W\ne \frac38$.

The orbifold compactification of the \EE\ heterotic string gives enough good phenomenologies, which is not competed in other superstrings. Yet, we have to resolve the moduli stabilization problem in this kind of heterotic string models.

\section{Acknowledgments}
Most results summarized in this talk have been collaborated with B. Kyae.
 This work is supported in part by the KRF Grants, No. R14-2003-012-01001-0 and No. KRF-2005-084-C00001.


\end{document}